\documentclass[aps,prd,preprint,superscriptaddress,nofootinbib,showpacs]{revtex4}
\usepackage{slashed}
\usepackage{epsfig,latexsym,cancel,amssymb,amsmath,verbatim,mathrsfs}
\usepackage{color}
\usepackage{graphicx}
\usepackage{ulem}
\usepackage{color}
\definecolor{My_red}        {cmyk}{0.00,1.00,1.00,0.20}

\def\ra{\rightarrow}
\def\L{\left(}
\def\R{\right)}

\def\f{\frac}
\newcommand{\be}{\begin{equation}}
\newcommand{\ee}{\end{equation}}
\newcommand{\bea}{\begin{eqnarray}}
\newcommand{\eea}{\end{eqnarray}}
\newcommand{\ba}{\begin{array}}
\newcommand{\ea}{\end{array}}

\long\def\symbolfootnote[#1]#2{\begingroup%
\def\thefootnote{\fnsymbol{footnote}}\footnote[#1]{#2}\endgroup}

\newcommand{\beq}{\begin{equation}}
\newcommand{\eeq}{\end{equation}}

\begin{document}

\title{Asymmetric Dark Matter Bound State}


\author{Xiao-Jun Bi}
\email[E-mail: ]{bixj@ihep.ac.cn}
\affiliation{Key Laboratory of Particle Astrophysics, Institute of High Energy Physics, Chinese Academy of Sciences, Beijing 100049, P. R. China}

\author{Zhaofeng Kang}
\email[E-mail: ]{zhaofengkang@gmail.com}
\affiliation{School of Physics, Korea Institute for Advanced Study,
Seoul 130-722, Korea}

\author{P. Ko}
\email[E-mail: ]{pko@kias.re.kr}
\affiliation{School of Physics, Korea Institute for Advanced Study,
Seoul 130-722, Korea}

\author{Jinmian Li}
\email[E-mail: ]{phyljm@gmail.com}
\affiliation{School of Physics, Korea Institute for Advanced Study,
Seoul 130-722, Korea}
\affiliation{ARC Centre of Excellence for Particle Physics at the Terascale, Department of Physics, University of Adelaide, Adelaide, SA 5005, Australia}

\author{Tianjun Li}
\email[E-mail: ]{tli@itp.ac.cn}
\affiliation{ Institute of Theoretical Physics, Chinese Academy of Sciences,
Beijing 100190, P. R. China }

\date{\today}

\begin{abstract}

We propose an interesting framework for asymmetric scalar dark matter (ADM),  which has novel collider phenomenology in terms of an unstable ADM bound state (ADMonium) produced via Higgs portals. ADMonium is a natural consequence of the basic features of ADM: the (complex scalar) ADM is charged under a dark local $U(1)_d$ symmetry which is broken at a low scale and provides a light gauge boson $X$. The dark gauge coupling is strong and then ADM can annihilate away into $X$-pair effectively. Therefore, the ADM can form bound state due to its large self-interaction via $X$ mediation. To explore the collider signature of ADMonium, we propose that ADM  has a two-Higgs doublet portal. The ADMonium can have a sizable mixing with the heavier Higgs boson, which admits a large cross section of ADMonium production associated with $b\bar b$. The resulting signature at the LHC depends on the decays of $X$. In this paper we consider a case of particular interest: $pp\ra b\bar b+ {\rm ADMonium}$ followed by ${\rm ADMonium}\ra 2X\ra 2e^+e^-$ where the electrons are identified as (un)converted photons. It may provide a competitive explanation to heavy di-photon resonance searches at the LHC.



\end{abstract}

\maketitle

\section{Introduction} 

Among a large number of models for dark matter (DM), ADM is one of the most attractive one because it provides a way to quantitatively relate the relic densities of dark and visible matters~\cite{Petraki:2013wwa,Zurek:2013wia}. Originally, it was proposed to understand their coincidence, namely $\Omega_{\rm DM}h^2:\Omega_{b}h^2\approx5:1$~\cite{Barr:1990ca}. More widely, ADM takes advantage in explaining DM relic density in a way insensitive to the strength of couplings involved in DM annihilating~\cite{Kang:2011ny}: It merely requires a sufficiently large cross section of DM-antiDM annihilating instead of a certain value.

Asides from the origin of { ADM} asymmetry that is beyond the scope of this paper, there are two ingredients for constructing an ADM model: A continuous symmetry $U(1)_d$ under which ADM is charged and a large DM-antiDM annihilation cross section to remove the symmetric part. A natural option is considering a dark sector which has a gauged $U(1)_d$ with a strong gauge coupling $g_d$. The actual dark matter number may not be gauged, but in the effective model  the gauged $U(1)_d$ at least guarantees that DM is not self-conjugate. Moreover, $U(1)_d$ spontaneously breaks at a low scale, giving rise to a light massive dark gauge boson $X$ into which ADM can annihilate. Additionally, the dark gauge interaction leads to ADM self-interaction, which may be good news for addressing the small scale problem; the $N-$body simulation of DM halos shows several discrepancies with the observations and they may be resolved by DM with large self-interaction~\cite{H. B. Yu}.~\footnote{ For a concrete example of local dark gauge symmetry for large self-interaction, see Ref.~ \cite{Ko:2014nha} where complex scalar dark matter with local $Z_3$ was discussed and compared with global $Z_3$ model. }

In this paper we point out that as a consequence of the above setup ADM can form bound state ADMonium, which may leave appreciable signals at the LHC, provided that ADM has sizable coupling to the Higgs sectors, says a 2HDM Higgs sector in this paper. ADMonium dominantly decays into a pair of $X$, so the concrete signal depends on the decay of $X$ into the standard model (SM) particles. We propose the fake di-photon signal, which arises when both the properly boosted $X$ { decay} into the displaced $e^+e^-$ pair. A detector simulation of such a phenomenon is furnished and model realization is provided, { based on the kinematic mixing between $X$ and gauge boson of the gauged lepton number $U(1)_\ell$.} Our study provides an alternative interpretation to the di-photon resonance searches at the LHC. Based on the latest data, we investigate the LHC sensitivity to the ADMonium behaving as a di-photon resonance.


The paper is organized as the following. In Section II we introduce the model. In Section III we present aspects of ADMonium. In Section IV we explore fake photon at LHC. Section V contains the discussion and conclusion.

\section{Asymmetric dark matter with a 2HDM Higgs portal} 


\subsection{The (effective) model setup}

In this paper we are interested in a scalar dark matter field $\chi$ which,  compared to the fermionic DM, possesses  special superiority in interactions with the visible sector  since it can always couple to the Higgs sector at the renormalizable level via the Higgs-portal term $|\chi|^2|\Phi_2|^2$ where $\Phi_2$ denotes the Higgs doublet in the SM. But here we would like to consider an extension to the SM Higgs sector by an extra Higgs doublet $\Phi_1$, {\it i.e.} the popular two-Higgs doublet models (2HDM). We will see that such an extension provides a new way to probe the dark sector at LHC (see other ways~\cite{Shepherd:2009sa,An:2015pva,Tsai:2015ugz}). For the purpose of LHC search, the ADM mass ($m_\chi$) is assumed to be $\sim{\cal O}(100)$ GeV, giving rise to a fairly heavy ADMonium; but a generation to other mass scale is straightfoward.~\footnote{{\noindent \bf{Light dark Higgs case}}
Before closing, let us make a comment on the light dark Higgs case. Let us denote the dark Higgs by $\phi_d$, which gives dark gauge boson $X$ mass after $U(1)_d$ symmetry breaking. If dark Higgs $h_d$ is light enough (say, $m_{h_d} \lesssim 2 m_e$ which may be rather a constrived condition ), it will have a large branching ratio for the diphoton final state. Then the dark Higgs portal $| \phi_d |^2 |\Phi_1|^2 $ term will make $h_1 \rightarrow h_d h_d$ which could be the main decay mode of $h_1$. This scenario itself could be interesting, but has nothing to do with AMDonium, and we do not consider further in this paper. 
}

To be specific, the effective interacting Lagrangian for the scalar ADM with a Higgs-portal is
\begin{eqnarray}\label{model}
-{\cal L}_{int}&=ig_d\chi^*\overleftrightarrow{\partial}_\mu   \chi X^\mu+g_d^2|\chi|^2X_\mu X^\mu+\eta_{i}|\chi|^2|\Phi_i|^2
+2\eta_{12}|\chi|^2{\rm Re}(\Phi_1^\dagger\Phi_2)+V(\Phi_1,\Phi_2),
\end{eqnarray}
We decompose the CP-even parts of $\Phi_i$ as ${\rm Re}(\Phi_i^0)=v_i+h_i/\sqrt{2}$ with $\sqrt{v_1^2+v_2^2}=174$ GeV. Through the kinetic mixing term, the $U(1)_d$ gauge boson $X_\mu$ slightly mixes with the gauge boson of $U(1)_\ell$, the { local lepton number} (We are working in the family universal case, but $\ell$ can be flavor dependent.), 
\begin{eqnarray}\label{kmixing}
-{\cal L}_{int}&=&-\f{1}{4}F_{X,\mu\nu}F_{X}^{\mu\nu}-\f{1}{4}F_{L,\mu\nu}F_\ell^{\mu\nu}-\f{\epsilon}{2}F_\ell^{\mu\nu}F_{X,\mu\nu}
+\f{m_X^2}{2}X_\mu X^\mu+\f{m_\ell^2}{2}L_\mu L^\mu,
\end{eqnarray}
with $m_X^2\ll m_\ell^2$. On the other hand, the mixing with the SM $U(1)_Y$ hypercharge gauge boson is totally negligible. We will explain the reason for such a setup later. After the {rotation} $L_\mu\ra L_\mu- \epsilon X_\mu$ and $X_\mu\ra 1/\sqrt{1+\epsilon^2} X_\mu$, we go to the canonical mass basis where the kinetic mixing term is eliminated. Eventually, $X$ couples to SM leptons, i.e., the lepton number current $J_\ell$~\cite{Baumgart}:
\begin{align}\label{X-decay}
-{\cal L}_X=\epsilon g_\ell  X_\mu J_\ell^\mu~~{\rm with}~~ J_\ell^\mu= Q_{L}^f\bar f_{L}\gamma^\mu  f_{L}+Q_{R}^f\bar f_{R}\gamma^\mu  f_{R},
\end{align}
where $Q_{L,R}^f$ is the lepton number of fermion $f_{L/R}$ and $g_\ell$ is the gauge coupling of $U(1)_\ell$, which is a free parameter and can be absorbed into $\epsilon$.

Comments on generalizations of the above setup are in orders. First, a fermionic ADM can be accommodated in the presence of a singlet Higgs field in the Higgs sector. Second, we choose a massive gauge boson as the force mediator, but a light Higgs boson can also play { the role}. {However,} the couplings of the scalar force mediator to SM particles are model dependent, for instance, to di-photon via a charged loop. Third, to avoid the Landau pole problem for $U(1)_d$, one may consider a non-Abelian dark gauge symmetry like $SU(N)$ which is completely broken down at a low scale {  and  behaves as a global charge for ADM. }

 \subsection{Bounds from DM direct detections and others}

Although ADM leaves no signals in indirect detections except for some special scenarios~\cite{stable:ADM,Feng:2013vva,Chen:2015yuz}, it leaves sufficient hints in direct detections in the scenario under consideration. As a matter of fact, the ADM-nucleon spin-independent (SI) scattering rate induced by the Higgs portal tends to exceed the upper bound set by the latest DM direct detection experiments such as LUX~\cite{Akerib:2016vxi} and PandaX-II~\cite{Tan:2016zwf}. On the other hand, the production of ${\cal D}$ and the ADM-nucleon SI scattering by exchanging $h_1$ are positively correlated via $(m_{h_1}, \eta_{12}, \tan\beta)$, so DM direct detections are able to impose stringent bounds on the fake di-photon rate from ADMonium.

ADM interacts with the visible sector via the 2HDM-portal and $X$-portal. For simplicity, we consider only the $\eta_{12}$ term among three Higgs portal terms. It generates both $h_{1,2}-\chi-\chi^*$ couplings, but the moderately heavier Higgs $h_1$, because of its $\tan^4\beta$ enhancement, always dominates over the SM-like Higgs boson $h_2$ in the large $\tan\beta$ limit. Then the DM-nucleon SI scattering cross section can be simply written as~\cite{Jungman:1995df,Gao:2011ka}
\begin{align}\label{}
\sigma_{\rm SI}^p=\f{m_p^2a_n^2}{\pi}~ {\rm with}~ a_n\approx0.22\f{m_n}{m_{\cal D}}\f{\eta_{12}}{m_{h_1}^2}\tan\beta O_{12}^2,
\end{align}
where the mixing factor $O_{12}\sim 1$ will be defined later. For a DM near 100 GeV, currently LUX yields the strongest upper bound on $\sigma_{\rm SI}^p\lesssim 10^{-10}$ pb~\cite{Akerib:2016vxi}. We parameterize the cross section to be
\begin{align}\label{h_1:SI}
\sigma_{\rm SI}^p=1.2\times10^{-9}\L\f{500\rm GeV}{m_{\cal D}}\R^6\L\f{\tan\beta\eta_{12}}{2}\R^2\L\f{O_{12}}{0.95}\R^4\rm pb,
\end{align}
where we have assumed close masses between $m_{h_1}$ and $m_{\cal D}$ so that the ADMonium-$h_1$ mixing angle can be enhanced. As one can see, a relatively light ADM does not admit a large $\tan\beta\eta_{12}\gg 1$ which can increase the production rate of $\cal D$; see Fig.~\ref{fig:lhc}.

At tree level, $X$ does not {mediate} ADM-nucleon SI scattering, which is the reason why we chose the kinematic mixing between $U(1)_d$ and $U(1)_\ell$ rather than $U(1)_Y$. Otherwise, the latter mixing will contribute to SI scattering with cross section 
\begin{align}\label{X:SI}
\sigma_{\rm SI}^{p,X}=& \f{m_p^2}{\pi}\f{g^2\sin^22\theta_w}{4}\f{g_d^2\epsilon_Y^2}{4}\f{1}{m_X^4}\equiv\f{m_p^2}{m_X^4}\hat\epsilon_Y^2\alpha_d\cr
=&4.0\times10^{-9}{\rm pb} \L\f{\hat\epsilon_Y}{10^{-8}}\R^2\L\f{1\rm GeV}{m_X}\R^4\f{\alpha_d}{0.1}.
\end{align}
This scattering rate is greatly enhanced by $1/m_X^4$, so $\hat\epsilon_Y$ should be extremely small to avoid direct detection exclusion. Consequently, we will see that it is impossible to make $X$, which decays too slowly,  mimic photon; see plots in Fig.~\ref{eff}. 

However, at loop level the charged leptons, which are charged both under $U(1)_\ell$ (thus under $U(1)_d$ with mini charge proportional to $\epsilon$) and QED, generate kinematic mixing between $X_\mu$ and photons. The strength is estimated to be~\cite{Fox:2008kb}
\begin{align}\label{}
\epsilon'\sim \f{e g_L \epsilon }{16\pi^2}\log\f{m_e}{m_\tau}.
\end{align}
Replacing $\hat \epsilon_Y$ with $\epsilon'$ in Eq.~(\ref{X:SI}), combining with Eq.~(\ref{length}), it is seen that $\epsilon'$ is near exclusion if we want to keep $P_X$ as high as possible; moreover, $m_X$ is favored to lie in the GeV scale. Note that for the $X$-mediating case, ADM interacts with nucleons in an isospin-violating manner~\cite{Gao:2011ka}, i.e., with proton only~\cite{Kang:2010mh}. This allows a substantial deconstructive interference effect between the two contributions Eq.~(\ref{h_1:SI}) and Eq.~(\ref{X:SI})~\cite{Gao:2011ka}, which may help to relax the bound. But that scenario is out the scope of this paper.

Laboratory searches for a light but leptonic dark gauge boson $X$ do not raise much concern for the typical strength of $g_L\epsilon\sim 10^{-6}$ and typical mass $m_X\sim {\cal O}(\rm GeV)$, which can be seen from a specific study~\cite{Jeong:2015bbi}. As a matter of fact, the strictest  constraint is imposed by stellar objects for $m_X\lesssim 1$ GeV. For instance, in supernova (SN) the $X_\mu$ mediated annihilation $e^+e^-\ra \nu_L\bar \nu_L$ with neutrinos escaping contributes to the cooling process of SN, which imposes a very strict upper bound on $g_L\epsilon $. If $g_L\epsilon $ becomes larger and leads to trapping of neutrinos inside the SN, the previous bound is invalid. This gives an additional bound which allows the region above that limit. In terms of the exclusion region shown in Ref.~\cite{Jeong:2015bbi}, we find that although not a definite exclusion, the SN constraint will be in strong tension with the dark photon scenario for $m_X\lesssim 1$GeV. Therefore, in this paper we will focus on the region $m_X\gtrsim1$ GeV.

\section{ADMonium}

In this section we will present the mechanism of ADM bound state formation, its mixing with Higgs bosons and its decays.

\subsection{Formation of ADMonium}

The force mediator $X$ leads to the formation of bound state for a 
pair of DM and antiDM with center-of-mass of energy near the threshold 
$m_{\cal D}\equiv 2m_\chi$; it is dubbed as ADMonium, unstable and distinguishable from the previous studies focusing on stable ADM bound state~\cite{stable:ADM}, e.g., the dark atom bounding two { different} species. 
To study the basic properties of ADMonium $\cal D$, the starting point 
is the attractive Yukawa potential (or the static screened Coulomb potential~\cite{Rogers}) 
between two ADM
\begin{align}\label{}
V(r)=-\f{\alpha_X}{r}e^{-m_X r},
\end{align}
with $\alpha_X>0$ for the DM and anti-DM system. However,  one has $\alpha_X<0$ namely a repulsive potential for the DM and DM or antiDM and antiDM system. If the self-interaction involves dimensionless couplings $g_d$ such as gauge coupling and Yukawa coupling, one has  $\alpha_X=g_d^2/4\pi$ from the single $X$ exchange diagram. For a scalar ADM $\chi$, DM self-interaction also arises from massive coupling like $A|\chi|^2 S$, with $S$ being some scalar (the SM Higgs boson or additional candidate) and then $\alpha_X=|A|^2/(16\pi m_\chi^2)$~\cite{Efimov:1999tx} (always positive). Perturbative bound requires $\alpha_X<1$; an even stronger upper bound will be derived.

{
Then one can solve the eigenvalues and corresponding eigenstates of the 
Schr$\ddot{\rm o}$dinger equation with this potential.}
The eigenstates are labelled by $(n,l)$ with $n$ and $l$ denoting the radial and the 
{ orbital} 
angular quantum numbers, respectively. In this paper we will focus on the case $l=0$,
i.e., the $ns$ state having wavefunction $\psi_n(\vec{x})=\f{1}{2\sqrt{\pi}}R_n(r)$ with $R_n(r)$ 
the radial wavefunction. For our purpose, the wavefunction at zero separation $\psi_n(0)$ is 
of interest. In the Coulomb limit, its square takes the form of
\begin{align}\label{}
|\psi_n(0)|^2=\f{1}{n^3}\f{1}{\pi a_0^3}=\f{1}{n^3}\f{\alpha_X^3 m_{\cal D}^3}{64\pi},
\end{align}
where $a_0=2/(\alpha_X m_\chi)$ is the Bohr radius of the $\chi$ pair system; it is the mean size of the ground state 1$s$. The mass of the $ns$ state is $m_{{\cal D}_n}=m_{\cal D}-E_n$ with $E_n$ the binding energy
\begin{align}\label{}
E_n=\f{\epsilon_n}{8}m_{\cal D}\alpha_X^2 ~\underrightarrow{D\gg1}~ \f{1}{8n^2}m_{\cal D}\alpha_X^2,
\end{align}
where $D=m_X^{-1}/a_0$ measures how the $\chi$ pair system is Coulomb-like. $D\gg1$ is the Coulomb limit where the range of the interaction is much larger than the typical size of the 1$s$ state. If $D$ is close to 1, $\epsilon_n$ will be much suppressed compared to the asymptotic value $1/n^2$; for instance, for $1s$ it reduces to merely 0.02 as $D=1$~\cite{Rogers}, resulting in a loose bound state. For $\alpha_X\ll1$, the mass splittings among different radial exciting states are negligible. Even though $\alpha_X^2\sim 0.1$, the widest mass splitting is still just $\lesssim0.01m_{\cal D}$.

There are a few conditions for the existence of at least one bound state, $1s$. In the first,  $1/m_X$, the screening length that characterizes the range of the interaction, should be at least longer than the Bohr radius; more concretely, one requires $D\gtrsim 0.84$~\cite{Rogers}. Immediately, we have $m_X\lesssim \alpha_X m_\chi/1.68$ and thus the mediator should be lighter than $\chi$. Second, the lifetime of ADMonium $ns$ should be longer than the time for ADMonium formation, i.e., the decay width $\Gamma_{\cal D}$ is smaller than the corresponding binding energy~\cite{Hagiwara}:
\begin{align}\label{}
2\Gamma_{\cal D}<E_n \ll \f{m_{\cal D}}{2},
\end{align}
where the second inequality is for the sake of reliability of non-relativistic approximation which always holds for $\alpha_X<1$. After using Eq.~(\ref{gamma:XX}) one can see that the first inequality imposes an upper bound on self-interaction coupling: $\alpha_X<(\epsilon_n/4)^{1/3}\simeq 0.6$ for $\epsilon_1=1$. A smaller $D$ yields a smaller upper bound on $\alpha_X$, for instance, 0.17 for $D=1$. In this paper { we consider the case where} 
$D\gg1$ holds.

\subsection{ADMonium-Higgs mixing}

To study the collider phenomenology of ADMonium, we should figure out its production and decay. At the LHC, a pair of free dark states can be produced and then they have certain probability to bound together 
{ near the threshold $m_{\cal D}$}. 
We consider the production mechanism of ADMonium by virtue of its mixing with 
the SM Higgs field after electroweak symmetry breaking.

Let us begin with a most general pattern of mixing. In the basis $\Phi^T=(h_1, h_2, {\cal D})$, the three by three { mass$^2$ matrix for three scalar bosons} takes the form of
\begin{align}\label{}
M_\phi^2=\left(\begin{array}{ccccc} m_{11}^2 \quad & m_{12}^2 \quad & \delta m^2_{1{\cal D}}
 \\  & m_{22}^2 \quad  & \delta m_{2\cal D}^2
  \\
  & &  m_{\cal D}^2-i\Gamma_{\cal D} m_{\cal D}
\end{array}\right),
\end{align}
a symmetric matrix. We have included the width of ADMonium before mixing, which may be relevant in the case of extremely degenerate between darkonimum and Higgs bosons. The Higgs bosons are always assumed to be  narrow resonances. The states in the mass eigenstates are labelled as $H^T=(H_3, H_2, H_1)$ with masses   
in descending order; they are related to $\Phi$ by the orthogonal matrix $O$: $\Phi=O H$.

The first two by two block of $M_\phi^2$, determined by the routine procedure in dealing with the Higgs potential, is not our focus. We focus on the off-diagonal elements, which are given by (with $v_{1,2}$ defined below Eq.~(\ref{model}))
\begin{align}\label{DHmixing}
& \delta m^2_{1(2){\cal D}}=2\f{|\psi(0)|}{\sqrt{m_{\cal D}}}\L\eta_{1(2)} v_{1(2)}
+\eta_{12} v_{2(1)}\R,
 \end{align}
where we focus on the $1s$ state of ADMonium since the production of the excited state is suppressed by $1/n^3$. To determine these mixing terms, we first calculate the bound state production from $gg\ra \chi\chi^*$ mediated by a Higgs boson, using the conventional way; then we instead use the mixing formalism and they should give the identical results. In this way one can gain the above expression. 

It is illustrative to consider the limits where only one Higgs-portal matters. We first consider the usual case, the { (SM Higgs doublet)} $\Phi_2$-portal. The mixing angle is given by
\begin{align}\label{mixing:2} 
\sin\theta_{2{\cal D}}
    \approx \f{\alpha_X^{3/2}}{4\sqrt{\pi}}\f{\eta_{2} v m_{\cal D}}{m_{\cal D}^2-m_{h_2}^2-i\Gamma_{\cal D}m_{\cal D}}.
     \end{align}
For an ADMonium much heavier than $m_{h_2}$, one has $\sin\theta\propto v/m_{\cal D}\ll1$; in the opposite, the mixing will be suppressed by $m_{\cal D}/m_{h_2}$. For safety, one may need to consider the region $m_\chi>m_{h_2}/2$. Typically, a fairly large $\alpha_X$ close to the perturbative limit is required to lift the mixing angle. Nevertheless, for $m_{h_2}\simeq m_{\cal D}$, it still can be of order 0.1 even for a relatively small $\alpha_X\sim 0.1$.

Now we move to the more interesting case in this paper, the $\Phi_1$-portal. The type-II 2HDM has the feature that  the associated production $h_1b\bar b$ is enhanced by large $\tan\beta=v_2/v_1$. Consequently, once ADMonium strongly mixes with $h_1$, the process $pp\ra {\cal D}b\bar b$ will be a promising way to produce ${\cal D}$. The $h_1 - {\cal D}$ mixing originates from the $\eta_{12}$-term and we can get the expression of mixing angle similar to Eq.~(\ref{mixing:2}). Again, substantial mixing happens only in the presence of more or less degeneracy between $h_1$ and ${\cal D}$:
\begin{align}\label{mixing:1} 
\sin\theta_{1{\cal D}}&\approx 0.15\L\f{\alpha_X}{0.3}\R^{\f{3}{2}}\L\f{\eta_{12}}{1.0}\R\f{({500\rm GeV})^2-({480\rm GeV})^2}{m_{\cal D}^2-m_{h_1}^2}.
\end{align}
The current LHC searches for extra heavy Higgs bosons decaying into $\tau \tau$ in the type-II 2HDM imposes a stringent constraint on the $m_{h_1}-\tan\beta$ plane~\cite{CMS:HA}; for instance, for $m_{h_1}=500$ GeV, $\tan\beta\gtrsim 30$ has been excluded at the 2$\sigma$ level. However, in our paper this constraint can be relaxed. The reason is that $h_1$ may decay into the dark sector particles such as dark Higgs boson, by which  Br$(h_1\ra\tau^+\tau^-)$ can be lowered substantially. Thus in this paper we will not incorporate this constraint.



\subsection{ADMonium annihilate decay}


Although the constitute is sufficiently stable, the bound state can annihilate decay. In general, the width of ${\cal D}\ra X_1X_2$ is formulated to be~\cite{Kats:2009bv}
\begin{eqnarray}\label{master}
\Gamma_{{\cal D}}(X_1X_2)&=&\f{1}{2m_{\cal D}}\f{1}{1+\delta_{AB}}\int d\Pi_2\f{2}{m_{\cal D}} |{\cal M}_{anni}|^2 |\psi(0)|^2\cr
&\equiv&\f{1}{1+\delta_{AB}}\f{ |\psi(0)|^2}{m_{\cal D}^2}W_{anni}(s),
\end{eqnarray}
with $\delta_{AB}$ the statistic factor. Thus, ADMonium decay is related with the annihilation of the free DM pair.

ADMonium decays in three ways: ({\it i}) Into a pair of force mediators,  which typically is dominant because of a large $\alpha_X$. The decay width is estimated as follows~\footnote{The corresponding cross section of DM-antiDM annihilation is huge: $\langle \sigma v\rangle_{XX}\approx \f{8\pi \alpha_X^2}{m_{\cal D}^2}\approx1.6\times10^3\times\L\f{\alpha_X}{0.3}\R^2\L\f{750{\rm GeV}}{m_{\cal D}}\R^2{\rm pb}.$ }
\begin{align}\label{gamma:XX}
\Gamma_{{\cal D}}(XX)&
\approx0.17\L\f{\alpha_X}{0.3}\R^5\L\f{m_{\cal D}}{500\rm GeV}\R\rm GeV~,
\end{align}
taking a massless $X$. ({\it ii}) Into the SM particles through its $h_{1,2}$ composition, but they are suppressed by either mixing or Yukawa couplings; ({\it iii}) Into the Higgs boson pairs via the Higgs-portal terms, and they are sizable only if the decay is kinematically accessible. In summary, it is reasonable to take ${\rm Br}_{\cal D}(XX)\approx100\%$.

\section{ ADMonium with fake di-photon signal } 

Heavy di-photon resonances appear in a lot of {models beyond SM}, in particular those involving extended Higgs sectors. At the same time, they can be well searched at the LHC due to the suppressed backgrounds. Therefore, they are the centeral topics both in the CMS and ATLAS collaborations. However, {an experimentally observed di-photon may not imply a resonance with two-body decaying into a pair of photon}~\cite{Agrawal:2015dbf,Knapen,Cho:2015nxy}: A light and boosted intermediate particle which decays into a pair of collimating $e^+e^-$ pair, says $X$ in this paper, can be misidentified as a  (un)converted photon~\cite{ATLAS-PHYS-PUB-2011-007, Khachatryan:2015iwa}. Thus, we can apply results from di-photon searches to see the prospect of dark sector specified by Eq.~(\ref{model}) and Eq.~(\ref{X-decay}). Additionally, our study may provide an alternative candidate for possible di-photon excess in the future.

\subsection{Di-photon signal without photons}

\begin{figure}[htb]
\begin{center}
\includegraphics[width=0.5\textwidth]{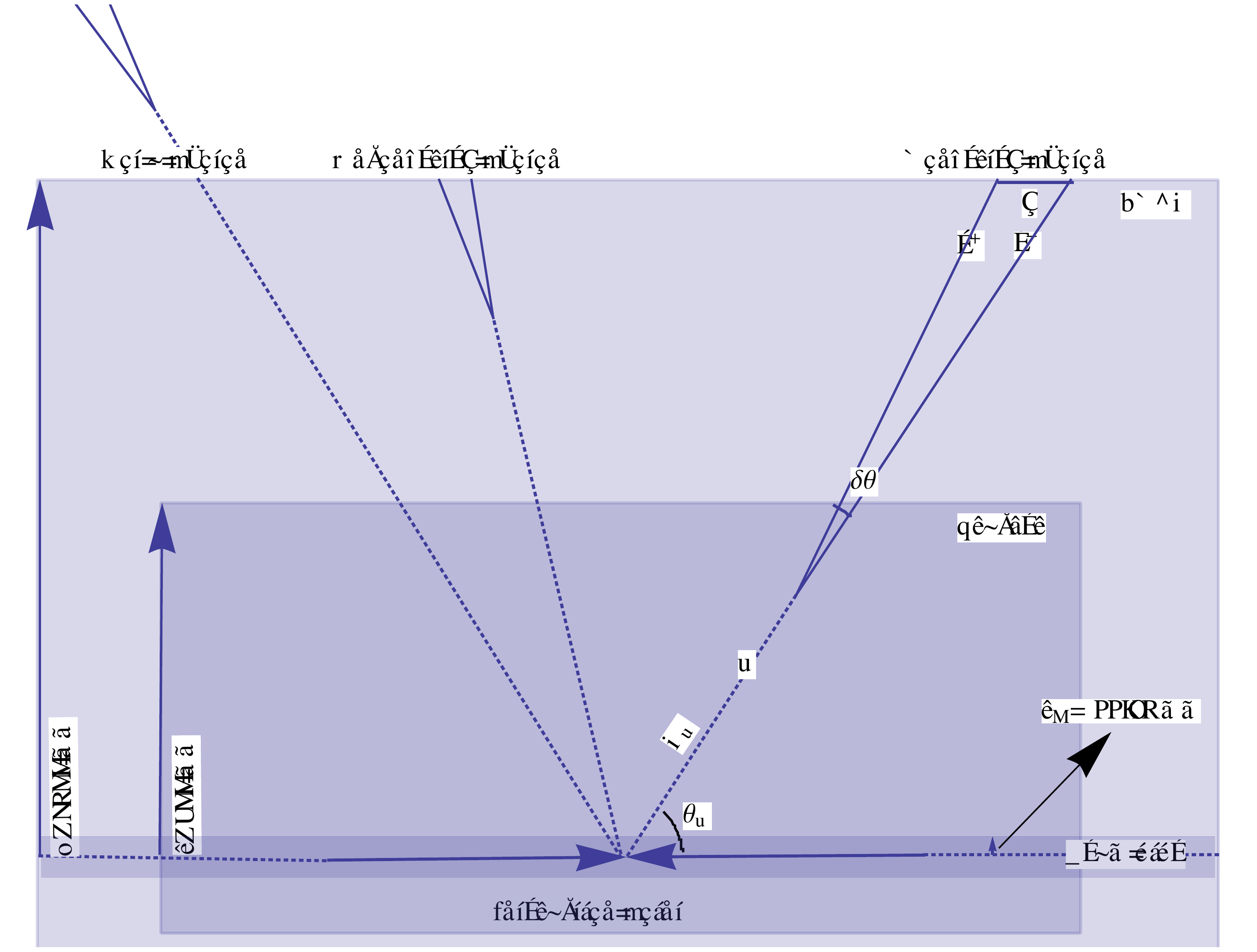}
\end{center}
\caption{The $e^+e^-$ from a light whilst energetic $X$ are aligned with each other and are able to mimic the photon conversion.}
\label{detector}
\end{figure}

Let us explain the emergence of a fake photon from $X \to e^+ e^-$ with more details. The main idea is illustrated in the schematic diagram Fig.~\ref{detector}.  {Several conditions are in orders: (1) Because photons can only be converted in the tracker when interacting with material, the flying $X$ should at least reach the pixel detector before decaying away, 
i.e., we should require $L_X  \sin \theta_X > 33.25$ mm~\cite{ATL-PHYS-PROC-2015-037} 
with $\theta_X$ and $L_X$ the polar angle and decay length of $X$, respectively; (2) %
To mimic a converted photon, $X$ should decay within the radius of 800 mm~\cite{Aad:2009wy} away from the interaction point (to have sufficient number of hits on the tracker) but within the radius of  [800 mm, $R$] with $R=1500$ mm the size of electromagnetic calorimeter (ECAL), in order to mimic an unconverted photon; (3) The $e^+$ and $e^-$ from the massless photon conversion have a small separation angle, as means that their distance at the ECAL layer is very small. Hence, conservatively we require that the reconstructed momentums of $e^+$ and $e^-$ from $X$ decay pointing to the same cell in ECAL. From Fig.~\ref{detector} one can calculate their distance
\begin{align}\label{d:es}
d\approx \L{R}/{ \sin \theta_X}  - L_X\R \delta \theta,
\end{align}
where $\delta \theta\sim m_X/p_{T,X}$ is the angular separation between two electrons. Taking the  ATLAS detector parameter for illustration, $0<d \lesssim 37.5$ mm {(corresponds to $\Delta \eta = 0.025$ granularity of the ECAL~\cite{Aad:2009wy})} is required. 

}

\begin{figure}[htb]
\includegraphics[width=2.8in]{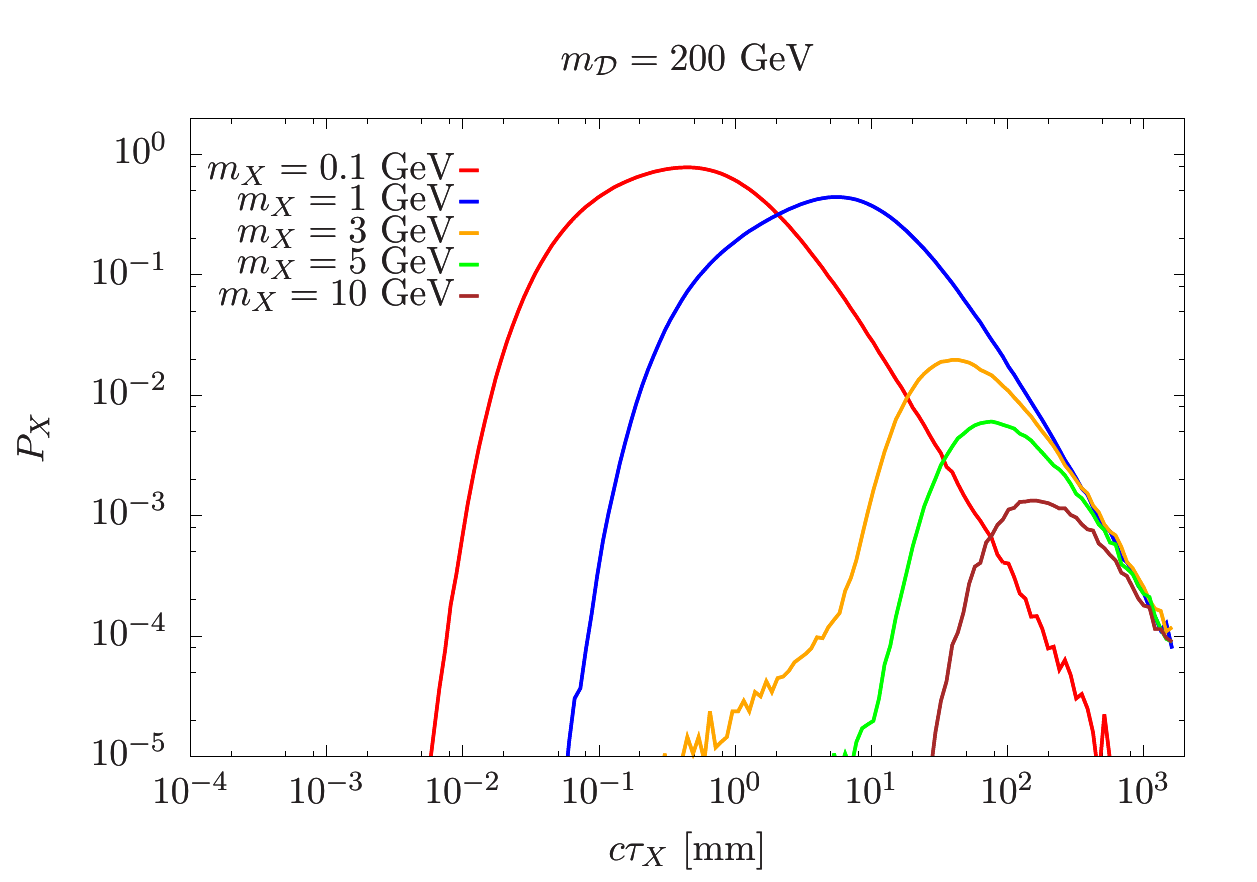}
\includegraphics[width=2.8in]{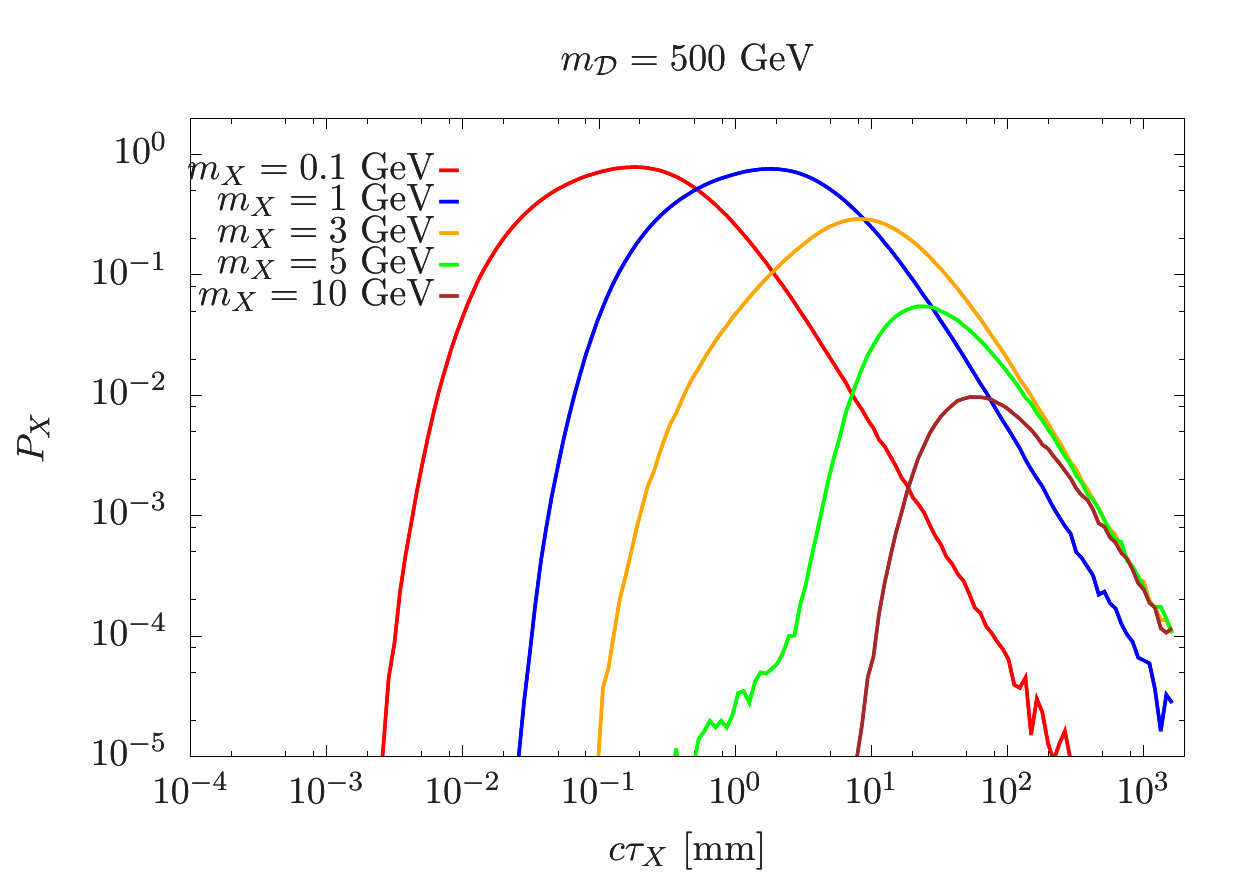}
\includegraphics[width=2.8in]{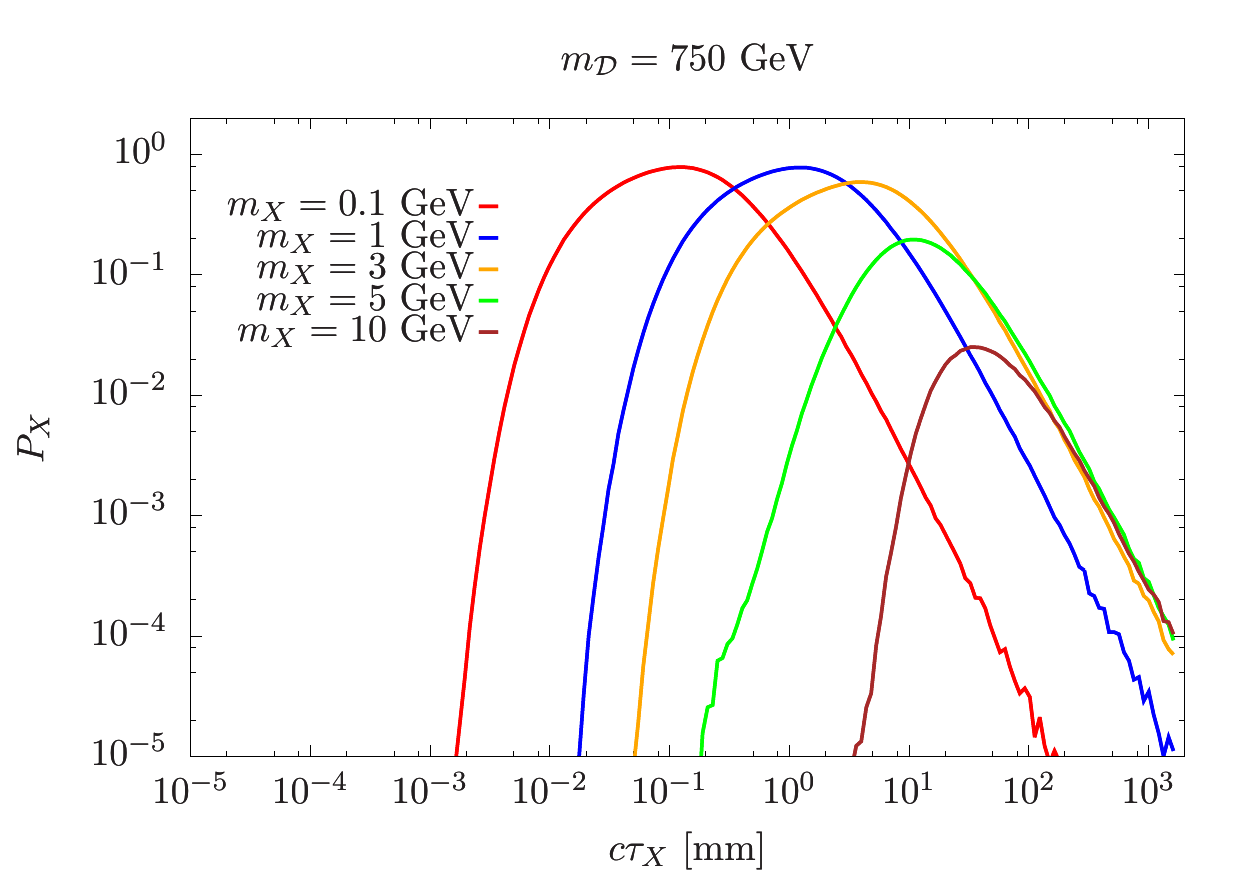}
\includegraphics[width=2.8in]{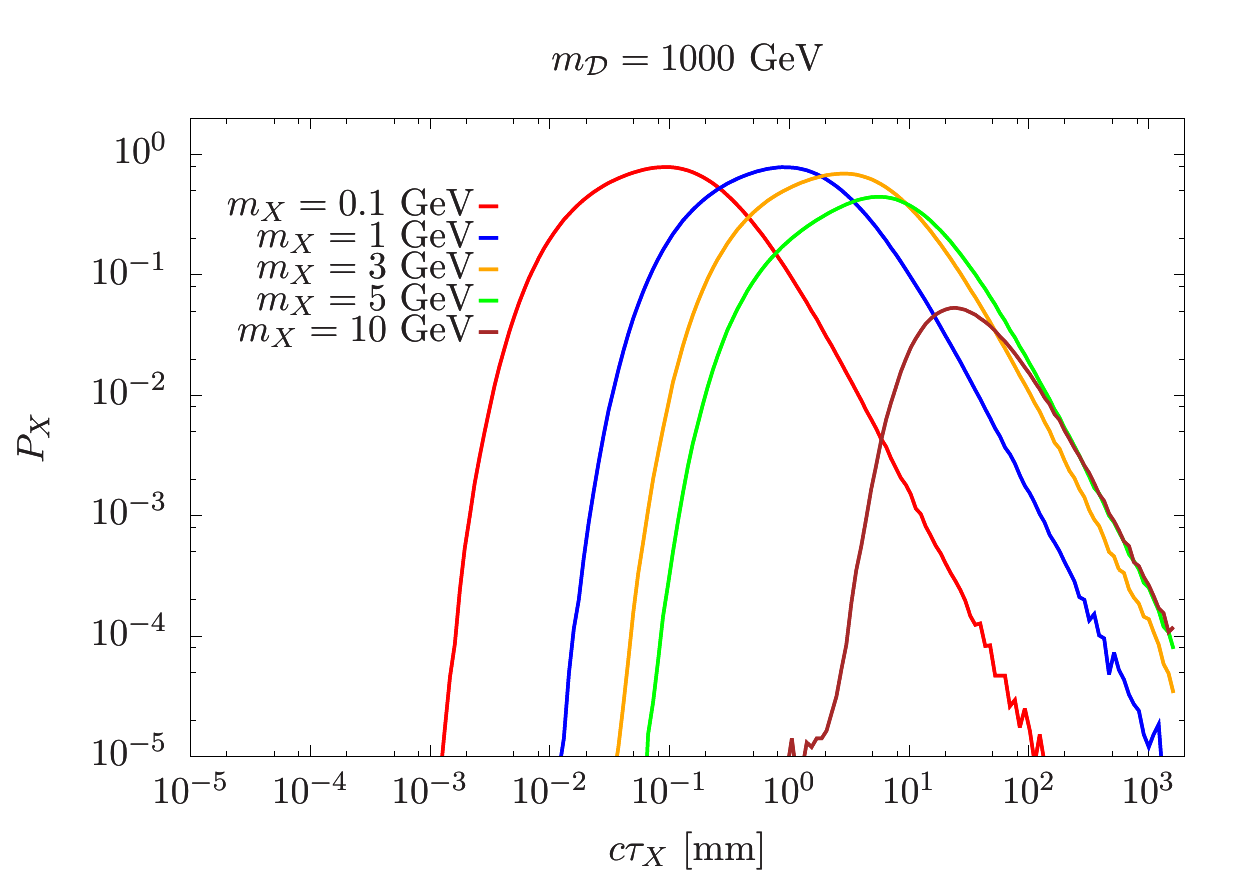}
\caption{The probability of di-$X(\ra e^+e^-)$ mimicing di-photon varies with the proper decay length of $X$ for $m_{\cal D} =(200, 500, 750, 1000)$ GeV. }\label{eff}
\end{figure}

To estimate $P_X$, the probability of the di-$X$ from $\cal D$ decay being identified as di-photon, we generate $10^6$ events of $gg \to {\cal D} (\to XX) b \bar{b}$ process with MadGraph5\_aMC@NLO~\cite{Alwall:2014hca} and count the number of events from which both $e^+e^-$ pairs satisfy the condition $L_X  \sin \theta_X > 33.25$ mm and $0<d \lesssim 37.5$ mm. For a given $m_{\cal D}$, in Fig.~\ref{eff} we show $P_X$ varying with the proper decay length of $X$, $c\tau_X$; a few values of $m_X$ are demonstrated. We see that $P_X$ peaks at certain $c\tau_X$ for a given $m_X$; as long as $X$ is sufficiently light, $P_X$ can be a few ten percents ($\lesssim75\%$) within a wide interval for $c\tau_X$. Note that to fit data, the required production cross section of $X$ pair is $\sim5/P_X$ fb. 

In a large parameter space of our dark sector $X$ behaves as a fake photon. The main decay of  the force mediator $X$ is via its coupling to $J_{L}$, having width
\begin{align}
\Gamma_X(f\bar f)=\f{ \epsilon^2g_\ell^2}{24\pi}\L(Q_{L}^f)^2+Q_{R}^f)^2\R m_X\L1+\f{2m_f^2}{m_X^2}\R\sqrt{1-\f{4m_f^2}{m_X^2}}.
\end{align}
For the decay into three flavor of active (almost massless) Majorana neutrinos, the total width is $\Gamma_X(\nu_L \nu_L)=3{ \epsilon^2g_\ell^2}m_X/{48\pi}$. Therefore, the proper decay length of $X$ is 
\begin{align}\label{length}
 c \tau_X\approx  14.5\times\L\f{10^{-6}}{\epsilon g_\ell}\R^2\L\f{\rm 1 GeV}{m_X}\R \rm mm.
\end{align}
In terms of the previous discussion, we should choose a relatively heavy $m_X\sim\rm GeV$ to avoid DM direct detection exclusions. In the laboratory frame, the proper decay length is enhanced by a large boost factor $\gamma_X=m_{\cal D}/2m_X$ and one then gets $L_X\approx  \gamma_Xc \tau_X$ which has been used in Eq.~(\ref{d:es}). The branching ratio of $e^+e^-$ mode is ${\rm Br}_X(e^+e^-)\approx 57\%$ and $36\%$ for $2m_\mu >m_X\gg 2m_e$ and $2m_\tau >m_X\gg 2m_\mu$, respectively. In summary, in the optimal case, the probability of ${\cal D}$ being hunted as a di-photon resonance can be as large as $P_X {\rm Br}{^2}_X(e^+e^-)\sim {\cal O}(10\%)$.

\subsection{ LHC sensitivity} 

\begin{figure}[thb] \centering
\includegraphics[width=0.47\textwidth]{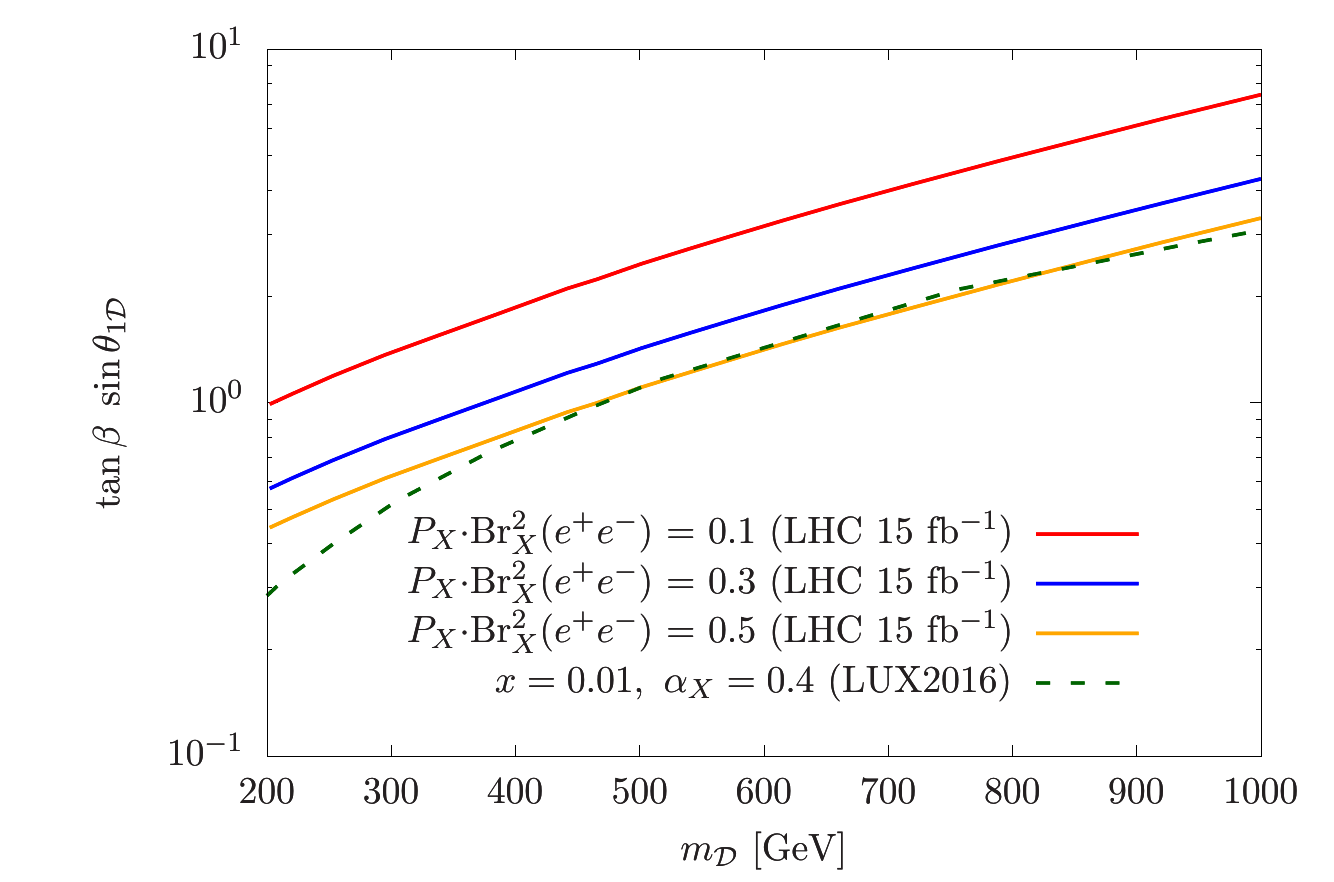}
\includegraphics[width=0.47\textwidth]{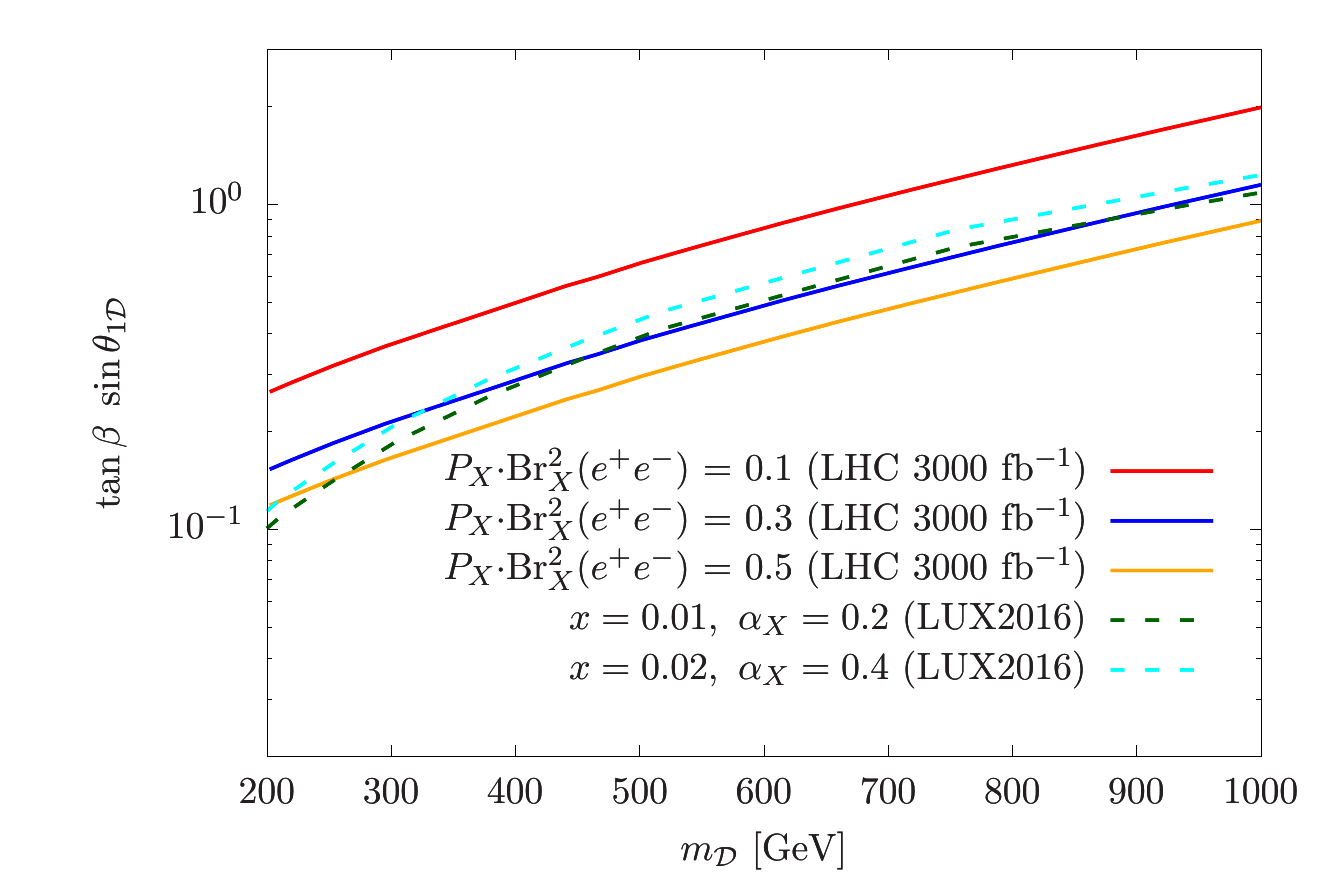}
 \caption{\label{fig:lhc} The 95\% CL exclusion limits from LHC search verse DM direct detecition at LUX~\cite{Akerib:2016vxi}. Left: di-photon search at current stage. Right: future prospect. }
\end{figure}

Now we investigate the LHC di-photon search sensitivities to ADMonium, from the latest 15/fb of $\sqrt{s}=13$ TeV ATLAS data~\cite{ATLAS:2016eeo} and the future prospect with higher luminosity calculated by using a simple extrapolation. The di-photon cross section $\sigma({\cal D}\ra 2X(\ra e^+e^-))$ mainly depends on five parameters, the ADMonium mass $m_{\cal D}$, ${\cal D}-h_1$ mixing angle $\theta_{1{\cal D}}$, $\tan\beta$ and ${\rm Br}_{X}(e^+e^-)$, $P_X$. But as mentioned before $P_X{\rm Br}^2_{X}(e^+e^-)$ can be fixed to its optimal value, and for concreteness 10\%, 30\% and 50\% will be chosen.~\footnote{The latter two values may be hard to achieve for the dark sector in our model, but we cannot exclude models where Br$(X\ra e^+e^-)\simeq100\%$, says those giving a light Higgs only coupling to electrons. } Then, we can demonstrate the current (left panel) and future (right panel) LHC sensitivity on the $m_{\cal D}-\tan\beta\sin \theta_{1{\cal D}}$ plane { in Fig.~\ref{fig:lhc}}. To show the stringent DM direct detection constraint, on the same plane we also add the upper bound (dashed lines) set by the complete LUX exposure~\cite{Akerib:2016vxi}, for which we take several samples of gauge coupling $\alpha_X$ and degeneracy $x$, defined through $m_{\cal D}=(1+x)m_{h_2}$; 
their concrete values are labeled in the { legends}. It is seen that, largely speaking,  the current LHC data is not { sensitive} to the parameter space allowed by LUX (below the dashed lines), except that one has even smaller $x$ or/and larger $\alpha_X$; while the increasing luminosity makes the region $\tan\beta\sin\theta_{1\cal D}\sim{\cal O}(0.1)$ for $m_{\cal D}\sim500$ GeV detectable, where the current DM direct detection experiments { can be evaded}. In addition to that, there may be ADM models that do not give rise to DM-nucleon scattering and then dashed lines can be removed.



To end up this section we would like to add a comment on the role of the associated $b$-jets. In our way to explain the di-photon data, two bottom quarks are accompanied at parton level, typically with only one of them is probable at detector. But moderate jet activity is allowed in the di-photon searches. To show this, we give an improper example. During the period of crazy ``750 GeV di-photon anomaly'', such a topic has been specifically discussed in Ref.~\cite{Harland-Lang:2016vzm}  based on the ATLAS measurement~\cite{Aaboud:2016tru}. The authors found that the $\chi^2$ of jet multiplicity distribution is 3.9 (5 Degree of Freedom) for the $bb$ initiated signal. However, the observed number of b-tagged events ($\sim$2.4 events) is around 2.4-$\sigma$ away from the prediction ($\sim$10 events). Thus there is a certain tension between the observation and the expectation. On the other hand, we would like to suggest the experimental searches to keep an eye on such kind of (heavy) jets associated with the di-photon, which is helpful to discriminate the production mechanisms of the resonance.

\section{Conclusions} 

ADM naturally forms bound state of DM as ADMonium. We established an effective model for ADMonium via 2HDM Higgs-portal and studied the LHC phenomenology. In this paper we consider a signal of particular interest: $pp\ra b\bar b+ {\rm ADMonium}$ followed by ${\rm ADMonium}\ra 2X\ra 2e^+e^-$ where the electrons are identified as (un)converted photons, which may provide a competitive explanation to heavy di-photon resonance searches.

\noindent {\bf{Note added:}} Soon after the submission of this paper, Refs.~\cite{Tsai:2016lfg,Chen:2016sck}  appeared on arxiv. Authors there used displaced $X\ra e^+e^-$ to explain the 750 di-photon excess, using slightly different parameters of the detector, but their results match ours well in the low $X$ mass region.

\noindent {\bf{Acknowledgements}} We would like to thank Guohuai Zhu, Hwidong Yoo and Pedro Schwaller for helpful discussions. This research is supported in part by: the Natural Science Foundation of China grant numbers 11475191, 11135009 (X.B), 11135003, 11275246, and 11475238 (T.L); the National Research Foundation of Korea (NRF) Research Grant NRF-2015R1A2A1A05001869, and by SRC program of NRF Grant No. 20120001176 funded by MEST through Korea Neutrino Research Center (KNRC) at Seoul National University (PK).


\vspace{-.3cm}
\begin{thebibliography}{99}




\bibitem{Petraki:2013wwa} 
  K.~Petraki and R.~R.~Volkas,
  Int.\ J.\ Mod.\ Phys.\ A {\bf 28}, 1330028 (2013). 

\bibitem{Zurek:2013wia} 
  K.~M.~Zurek,
  Phys.\ Rept.\  {\bf 537}, 91 (2014). 

\bibitem{Barr:1990ca}
  S.~M.~Barr, R.~S.~Chivukula and E.~Farhi,
  Phys.\ Lett.\  B {\bf 241}, 387 (1990);
  D.~B.~Kaplan,
  Phys.\ Rev.\ Lett.\  {\bf 68}, 741 (1992);
  N.~Cosme, L.~Lopez Honorez and M.~H.~G.~Tytgat,
  Phys.\ Rev.\  D {\bf 72}, 043505 (2005);
  R.~Kitano, H.~Murayama and M.~Ratz,
  Phys.\ Lett.\  B {\bf 669}, 145 (2008);
  H.~An, S.~L.~Chen, R.~N.~Mohapatra and Y.~Zhang,
  JHEP {\bf 1003}, 124 (2010); 
  D.~E.~Kaplan, M.~A.~Luty and K.~M.~Zurek,
  Phys.\ Rev.\  D {\bf 79}, 115016 (2009); 
  P. H. Gu, M. Lindner, U. Sarkar and X. Zhang, arXiv:1009.2690; 
  P.~S.~B.~Dev and D.~Teresi,
  arXiv:1512.07243 [hep-ph].

\bibitem{Kang:2011ny} 
  Z.~Kang and T.~Li,
  JHEP {\bf 1210}, 150 (2012). 


\bibitem{H. B. Yu} 
D. N. Spergel and P. J. Steinhardt, Phys. Rev. Lett. 84, 3760 (2000); S. Tulin, H. B. Yu and K. M. Zurek, Phys. Rev. D 87, no. 11, 115007 (2013).

\bibitem{Ko:2014nha} 
  P.~Ko and Y.~Tang,
  JCAP {\bf 1405}, 047 (2014).

\bibitem{An:2015pva}
  H.~An, B.~Echenard, M.~Pospelov and Y.~Zhang,
  arXiv:1510.05020 [hep-ph].


\bibitem{Tsai:2015ugz}
  Y.~Tsai, L.~T.~Wang and Y.~Zhao,
  arXiv:1511.07433. 

\bibitem{Shepherd:2009sa}
  W.~Shepherd, T.~M.~P.~Tait and G.~Zaharijas,
  Phys.\ Rev.\ D {\bf 79}, 055022 (2009).



 \bibitem{Baumgart} 
M. Baumgart, C. Cheung, J.T. Ruderman, L.-T. Wang and I. Yavin, 
JHEP 04 (2009) 014. 

 

\bibitem{stable:ADM} 
 D.~E.~Kaplan, G.~Z.~Krnjaic, K.~R.~Rehermann and C.~M.~Wells,
  JCAP {\bf 1005}, 021 (2010); 
D. E. Kaplan, G. Z. Krnjaic, K. R. Rehermann, and C. M. Wells, 
JCAP 1110 (2011) 011; J. M. Cline, Z. Liu, G. Moore, and W. Xue, 
Phys.Rev. D89 (2014) 043514; K. Petraki, L. Pearce, and A. Kusenko, 
JCAP 1407 (2014) 039; M. B. Wise and Y. Zhang, 
Phys.Rev. D90 (2014) 055030;   R.~Laha and E.~Braaten,
  Phys.\ Rev.\ D {\bf 89}, no. 10, 103510 (2014); 
  R.~Laha, 
  Phys.\ Rev.\ D {\bf 92}, 083509 (2015); 
K.~Petraki, M.~Postma and M.~Wiechers,
  JHEP {\bf 1506}, 128 (2015); 
  R.~Foot and S.~Vagnozzi,
  Phys.\ Rev.\ D {\bf 91}, 023512 (2015).

\bibitem{Chen:2015yuz} 
  S.~L.~Chen and Z.~Kang,
  arXiv:1512.08780 [hep-ph].

\bibitem{Feng:2013vva} 
  L.~Feng and Z.~Kang,
  JCAP {\bf 1310}, 008 (2013).

  \bibitem{Akerib:2016vxi} 
  D.~S.~Akerib {\it et al.},
  arXiv:1608.07648 [astro-ph.CO].

 \bibitem{Tan:2016zwf} 
  A.~Tan {\it et al.} [PandaX-II Collaboration],
  Phys.\ Rev.\ Lett.\  {\bf 117}, no. 12, 121303 (2016).


\bibitem{Jungman:1995df}
  G.~Jungman, M.~Kamionkowski and K.~Griest,
  Phys.\ Rept.\  {\bf 267}, 195 (1996).
  



\bibitem{Gao:2011ka} 
  X.~Gao, Z.~Kang and T.~Li,
  JCAP {\bf 1301}, 021 (2013).

\bibitem{Fox:2008kb} 
  P.~J.~Fox and E.~Poppitz,
  Phys.\ Rev.\ D {\bf 79}, 083528 (2009).



\bibitem{Kang:2010mh} 
  Z.~Kang, T.~Li, T.~Liu, C.~Tong and J.~M.~Yang,
  JCAP {\bf 1101}, 028 (2011).



\bibitem{Jeong:2015bbi} 
  Y.~S.~Jeong, C.~S.~Kim and H.~S.~Lee,
  Int.\ J.\ Mod.\ Phys.\ A {\bf 31}, no. 11, 1650059 (2016).

\bibitem{Rogers}
F. J. Rogers, H. C. Graboske, Jr., and D. J. Harwood, Phys. Rev. A 1, 1577 (1970).


\bibitem{Efimov:1999tx} 
  G.~V.~Efimov,
  hep-ph/9907483.

\bibitem{Hagiwara}
K. Hagiwara, K. Kato, A.D. Martin and C.?K. Ng, Nucl. Phys. B344, 1 (1990).


\bibitem{CMS:HA}
  The ATLAS collaboration,
  ``Search for Neutral Minimal Supersymmetric Standard Model Higgs Bosons $H/A \to \tau \tau$ produced in $pp$ collisions at $\sqrt{s}=13$ TeV with the ATLAS Detector,''
  ATLAS-CONF-2015-061.


\bibitem{Kats:2009bv}
  Y.~Kats and M.~D.~Schwartz,
  JHEP {\bf 1004}, 016 (2010).



\bibitem{Knapen}
S. Knapen, T. Melia, M. Papucci and K. Zurek, 
arXiv:1512.04928; M. Chala, M. Duerr, F. Kahlhoefer and K. Schmidt-Hoberg, 
arXiv:1512.06833;  
J. Chang, K. Cheung and C. T. Lu, arXiv:1512.06671;  X.~J.~Bi {\it et al.},
  arXiv:1512.08497; 
    L.~Aparicio, A.~Azatov, E.~Hardy and A.~Romanino,
  arXiv:1602.00949; 
  U.~Ellwanger and C.~Hugonie,
  arXiv:1602.03344 [hep-ph]; 
  B.~Dasgupta, J.~Kopp and P.~Schwaller,
  arXiv:1602.04692 [hep-ph].



\bibitem{Agrawal:2015dbf}
  P.~Agrawal, J.~Fan, B.~Heidenreich, M.~Reece and M.~Strassler,
  arXiv:1512.05775 [hep-ph].

\bibitem{Cho:2015nxy} 
  W.~S.~Cho, D.~Kim, K.~Kong, S.~H.~Lim, K.~T.~Matchev, J.~C.~Park and M.~Park,
  arXiv:1512.06824 [hep-ph].


\bibitem{Khachatryan:2015iwa} 
  V.~Khachatryan {\it et al.} [CMS Collaboration],
  JINST {\bf 10}, no. 08, P08010 (2015). 


\bibitem{ATLAS-PHYS-PUB-2011-007}
 ATLAS collaboration, 
 ATLAS-PHYS-PUB-2011-007

\bibitem{ATL-PHYS-PROC-2015-037}
 ATLAS collaboration, 
ATL-PHYS-PROC-2015-037



\bibitem{Aad:2009wy} 
  G.~Aad {\it et al.} [ATLAS Collaboration],
  arXiv:0901.0512. 

\bibitem{Alwall:2014hca} 
  J.~Alwall {\it et al.},
  JHEP {\bf 1407}, 079 (2014)
  doi:10.1007/JHEP07(2014)079
  [arXiv:1405.0301 [hep-ph]].

\bibitem{ATLAS:2016eeo} 
  The ATLAS collaboration [ATLAS Collaboration],
  ATLAS-CONF-2016-059.
  
  \bibitem{Harland-Lang:2016vzm} 
  L.~A.~Harland-Lang, V.~A.~Khoze, M.~G.~Ryskin and M.~Spannowsky,
  arXiv:1606.04902 [hep-ph].

\bibitem{Aaboud:2016tru} 
  M.~Aaboud {\it et al.} [ATLAS Collaboration],
  arXiv:1606.03833 [hep-ex].

 

\bibitem{Tsai:2016lfg} 
  Y.~Tsai, L.~T.~Wang and Y.~Zhao,
  arXiv:1603.00024 [hep-ph].
 
 \bibitem{Chen:2016sck} 
  C.~Y.~Chen, M.~Lefebvre, M.~Pospelov and Y.~M.~Zhong,
  arXiv:1603.01256 [hep-ph].






 


\end{thebibliography}
\end{document}